\def \CQ             {{1997~CQ$_{29}$}}

\def \WW             {{1998~WW$_{31}$}}

\def \TC             {{(47171)~1999~TC$_{36}$}}

\def \CM             {{(80806)~2000~CM$_{105}$}}

\def \CS             {{(79360)~1997~CS$_{29}$}}

\def \OJ             {{1999~OJ$_{4}$}}
\def \YW             {{(82075)~2000~YW$_{134}$}}

\def \OJS             {{2000~OJ$_{67}$}}

\def \TL             {{(48639)~1995~TL$_{8}$}}

\documentclass{iaus}
\usepackage{graphicx}

\title[Solar System Binaries] 
{Solar System Binaries}

\author[K.~S.~Noll]   
{Keith S.~Noll$^1$
}

\affiliation{$^1$Space Telescope Science Institute, 3700 San Martin Drive, Baltimore, MD 21218, USA \break email: noll@stsci.edu\\[\affilskip]}

\pubyear{2005}
\volume{229}  
\pagerange{1-20}
\date{7 August 2005}
\jname{Asteroids, Comets, Meteors 2005}
\editors{S. Ferraz-Mello \& D. Lazzaro, eds.}
\begin{document}

\maketitle

\begin{abstract}
The discovery of binaries in each of the major populations of minor bodies in the solar system is propelling a rapid growth of heretofore unattainable physical information. The availability of mass and density constraints for minor bodies opens the door to studies of internal structure, comparisons with meteorite samples, and correlations between bulk- physical and surface-spectral properties. The number of known binaries is now more than 70 and is growing rapidly. A smaller number have had the extensive followup observations needed to derive mass and albedo information, but this list is growing as well. It will soon be the case that we will know more about the physical parameters of objects in the Kuiper Belt than has been known about asteroids in the Main Belt for the last 200 years. Another important aspect of binaries is understanding the mechanisms that lead to their formation and survival. The relative sizes and separations of binaries in the different minor body populations point to more than one mechanism for forming bound pairs. Collisions appear to play a major role in the Main Belt. Rotational and/or tidal fission may be important in the Near Earth population. For the Kuiper Belt, capture in multi-body interactions may be the preferred formation mechanism. However, all of these conclusions remain tentative and limited by observational and theoretical incompleteness. Observational techniques for identifying binaries are equally varied. High angular resolution observations from space and from the ground are critical for detection of the relatively distant binaries in the Main Belt and the Kuiper Belt. Radar has been the most productive method for detection of Near Earth binaries. Lightcurve analysis is an independent technique that is capable of exploring phase space inaccessible to direct observations. Finally, spacecraft flybys have played a crucial paradigm-changing role with discoveries that unlocked this now-burgeoning field.

\keywords{Kuiper Belt, minor planets, asteroids, planets and satellites: formation}
\end{abstract}

\section{Two Are Better Than One}
From time to time in science there are paradigm-shifting developments that occur at such a rapid pace that it is recognized that an important new subfield of science has emerged.  Such is the case for the study of gravitationally bound companions to minor bodies in the solar system, a group of objects that can be referred to as {\em binary minor planets}, or, more simply, {\em binaries}.  The broad term {\em binary} used in this context often includes objects that would normally be considered satellites because of the large mass ratio of primary to secondary, others that are true binaries or doubles where the mass ratio of primary to secondary is closer to one, and contact binaries and bilobate objects.  Systems consisting of more than two gravitationally bound objects can also exist and one such system has now been observed (Marchis et al.\ 2005a).  

For the lingustic purists, it is worth noting that the term {\em binary} is frequently used to refer to gravitationally bound minor planets regardless of the mutual size of the components.  If one cared to do so, a {\em true} binary could be defined as a system where the barycenter resides outside of either of the two gravitationally bound bodies.  By this definition, Pluto and Charon qualify as the first known solar system binary (not counting the Sun-Jupiter binary).  The requirement for meeting this definition can be expressed as $$  a > {r_p (1+ {m_p\over m_s})} $$  where a is the semimajor axis of the system, $m_p$ and $m_s$ are the masses of the two components, and $r_p$ is the radius of the primary which can be expressed in terms of the primary mass and density as $r_p = ( 3m_p/4\pi \rho )^{1/3}$.   In practice, however, many systems will not have measured radii or known densities and albedos and thus the location of the barycenter relative to the surfaces will be uncertain.  As with many other instances of taxonomical terminology, however, the precise use of the term {\em binary} is less important than is a detailed knowledge of the objects in question.  In this review I will use the term {\em binary} in the broadest sense to include both true binaries and related classes of objects found among the minor body populations of the solar system.

\section{The Discovery of Solar System Binaries}

The two-century-long history of searches for and eventual discovery of binary minor planets has been summarized in several earlier reviews, most recently and most thoroughly by Merline et al.\ (2002) and by Richardson \& Walsh (2005).  Several milestones in that history are especially noteworthy and are summarized in Table~\ref{tab:hist}.  

\begin{table}\def~{\hphantom{0}}
  \begin{center}
  \caption{Significant Milestones in Solar System Binaries}
  \label{tab:hist}
  \begin{tabular}{llll}\hline
  Date   	&  Object			&  Significance    							&  Reference 		\\\hline
  1801   	&  (1) Ceres 		& search for asteroid satellites begins				&           					\\
  1901  	&  (433) Eros		& claim of double from lightcurve				& Andr\'e\ 1901				\\
  1978  	&  Pluto/Charon		& first binary TNO			  				& Christy \& Harrington\ 1978 	\\
  1990  	&  (4769) Castalia 	& first bilobate NEO imaged 					& Ostro et al.\ 1990 			\\
  1993	&  (243) Ida		& first Main Belt satellite						& Belton \& Carlson\ 1994 	\\
  1997	&  1994 AW$_1$	& first asynchronous lightcurve NEA				& Pravec \& Hahn\ 1997			\\
  1998	&  (45) Eugenia		& first groundbased detection of main belt satellite	& Merline et al.\ 1999a,b		\\
  2001	&  2000 DP$_{107}$	& first binary NEA radar detection				& Ostro et al.\ 2000			\\
  2001	&  1998 WW$_{31}$	& second binary TNO 						& Veillet et al.\ 2001 			\\
  2005	&  (87) Sylvia		& first multiple system						& Marchis et al.\ 2005a		\\\hline
  \end{tabular}
  \end{center}
  \end{table}

Over time there have been significant shifts in the prevailing view of the prevalence and nature of binaries in the solar system.  Almost immediately after the discovery of (1) Ceres by G.~Piazzi in early 1801, searches for satellites began.  These searches continued with varying degrees of intensity for nearly two centuries without the definitive detection of a satellite or binary.  Reports of possible binaries in the 20$^{th}$ century were numerous, the first being an analysis of the lightcurve of (433) Eros in comparison to the lightcurves of spectroscopic double stars (Andr\'e 1901).  The long series of unsuccessful searches had become so discouraging by the early 1980s that Weidenschilling et al.\ (1989) titled their review article ``Do Asteroids Have Satellites?".  It is interesting to note that by that time Pluto's binary companion Charon had already been discovered, but the existence of a large transneptunian population of small bodies was unknown and Pluto was still solidly ensconced as a major planet.  

The veil began to be drawn back in 1990 with the discovery of the bilobed nature of the NEA 1989 PB, now known as (4769) Castalia (Ostro et al.\ 1990).   The breakthrough came with the serendipitous discovery of (243) Ida's satellite Dactyl; unlike all previous reports this claim was indisputable.  While the reality of bound systems among the minor bodies of the solar system could no longer be questioned, it remained to be seen whether the Ida/Dactyl pair was an anomaly or representative of a significant, but as yet undetected population of bound systems.  In 1997-1998 this question was dramatically answered by two major steps forward:  the detection of the first asynchronous lightcurve binary (Pravec et al.\ 1997) and the first ground-based detection of an asteroid satellite (Merline et al.\ 1999a).  The method of identifying binaries by resolving complex lightcurves into two simple periodic components punctuated by mutual events pioneered by Pravec and colleagues was not universally accepted initially and so the first radar detection of a binary, 2000 DP$_{107}$ by Ostro et al.\ (2000) inaugurated an important new and definitive technique for the identification of NEA binaries.  Soon after, this object was also found to have eclipse/occultation events on top of the rotation lightcurve of the primary (Pravec et al.\ 2000a); an important step in proving the reliability of the lightcurve method that is responsible for a large fraction of the known NEA and Main Belt binaries (see Tables~\ref{tab:neb}-\ref{tab:mbb}).  The detection of the companion of \WW , which we now understand to be the {\em second} known binary in the TNO population shattered several prevailing assumptions, including the assumed uniqueness of the Pluto/Charon system and the thought that other transneptunian binaries, if they did exist, would have separations and orbits on a scale comparable to the Pluto/Charon system and would thus be difficult to detect around all but the largest TNOs, even with HST.  

Taken together, the change in little more than a decade since the discovery of Dactyl can only be described as breathtaking.  With such a rapid pace of development, the process of reviewing a field is daunting.  In the sections below I have tried to identify aspects of this field that are currently the best developed or where important new shifts in thinking are taking place.  This is not the first review of this field and will certainly not be the last.  It is my modest hope, however, that it is a fair snapshot of where we are at the time of the IAU Symposium 229, Asteroid, Comets, Meteors in 2005.

\section{The Current Inventory}

The number of known or suspected binary systems continues to grow rapidly.  Table~\ref{tab:neb} summarizes the inventory of binary minor planets reported as of October 2005.  The table includes objects directly observed with imaging instruments or radar and systems identified through lightcurve analysis.  Not included are reported instances of possible occultations by companions that are not generally accepted as sufficient evidence for a claim of a binary, although some could be genuine.  These more tenuous claims have been summarized by (Weidenschilling et al.\ 1989).  The total number of binary systems in Tables~\ref{tab:neb}-\ref{tab:tnb} is 73 including, 24 NEA binaries, 26 Main Belt binaries, 1 Jupiter Trojan, and 22 transneptunian binaries.  Despite searches, no binary Centaurs have been found.  Neither of the 2 Neptune Trojans are known to be binary.  There are insufficient statistics, however, to determine whether these non-detections are of any significance.  However, searches for binaries in these populations would certainly be worthwhile both for the potential physical information that binaries can yield and for studies of binary statistics in different populations.  With just a few tens of objects spread over many possible groupings of size, spectral type, dynamical class, etc., this list, though impressive, is still inadequate for many of the kinds of questions we would like to ask.  It is clear, however, that the known objects are only a tiny fraction of the population of bound systems that are potentially detectable with current and near-future technology.

\begin{table}\def~{\hphantom{0}}
  \begin{center}
  \caption{Near Earth Binaries}
  \label{tab:neb}
  \begin{tabular}{lllllr}\hline
	      	 			& \multispan2{semimajor~axis\ }&			& Period		& reference$^1$ 	\\ %
 Object      			& a (km) 		& a/r$_p$  	&  r$_s$/r$_p$	& (hours)		&				\\\hline
					&			&			&			&			&				\\ %
{\it radar$^2$}			&			&			&			&			&				\\ %
2000 DP$_{107}$		& 2.6(2)		& {\it 6.5}		& 0.41(2)		& 42.2(1)		& [Mg02][P05a]		\\ 
2000 UG$_{11}$		& {\it 0.3}		& {\it 2.6}		& {\it 0.36(9)}	& 18.4(1/2)	& [No00][P05a]		\\ %
(66391) 1999 KW$_4$	& 			& 			& {\it 0.3-0.4}	& 17.44(1)	& [Be01][PS01][P05a]	\\ %
1998 ST$_{27}$		& {\it $<$7}	&{\it $<$9}	& {\it 0.15}	& 			& [Be03]			\\ %
2002 BM$_{26}$		& 			& 			& {\it 0.2}		&{\it $<$72}	& [No02a]			\\ %
2002 KK$_8$			&			& 			& {\it 0.2}		&			& [No02b]			\\ %
(5381) Sekhmet			& 1.54(12)	& {\it 3.1}		& {\it 0.3}		& 12.5(3)		& [No03a][Ne03]	\\ %
2003 SS$_{84}$		&	 		&			& {\it 0.5}		& {\it 24}	 	& [No03b]			\\ %
(69230) Hermes			& {\it 0.6}		& {\it 2.5-4}	& {\it 0.9(1)}	&{\it 13.894(4)}& [Mg03][P05a]		\\ %
1990 OS				& {\it $>$ 0.6}	& {\it $>$4}	& {\it 0.15}	& {\it 18-24}	& [Os03]			\\ %
2003 YT$_1$			& {\it 2.7}		& {\it 5}		& 0.19(9)		& {\it 30}		& [No04a,b]		\\ %
2002 CE$_{26}$		& {\it 5}		& {\it 3}		& {\it 0.05}	& {\it 16}		& [Sh04][Sl04]		\\ %
					&			&			&			&			&				\\ %
{\it lightcurve$^2$}		&			&			&			&			&				\\ %
1994 AW$_1$			& 			& 			& 0.49(2)		& 22.33(1)	& [PH97][P05a]		\\ %
(35107) 1991 VH	 	& {\it 6.5(2.0)}	& 5.4(6)		& 0.37(3)		& 32.66(5)	& [P98][P05a]		\\ %
(3671) Dionysus			& 			& 4.5(1.0)		& 0.20(2)		& 27.74(1)	& [Mo97][P05a]		\\ %
1996 FG$_3$			& 			& 3.2(4)		& 0.29(2)		& 16.135(10)	& [P00b][ML00][P05a]	\\ %
(5407) 1992 AX			& 			& 			& 0.2(1)		&{\it13.520(1)}	& [P00b][P05a]		\\ %
(31345) 1998 PG		& 			& 			& {\it 0.3}		&{\it14.005(1)}	& [P00b][P05a]		\\ %
(88710) 2001 SL$_9$	 	& 			& 			& 0.28(2)		& 16.40(2)	& [P01a][P05a]		\\ %
1999 HF$_1$			& 			& 4.0(6)		& 0.22(3)		& 14.02(1)	& [P02][P05a]		\\ %
(66063) 1998 RO$_1$	& 			& 			& 0.48(3)		& 14.54(2)	& [P03a][P05a]		\\ %
(65803) Didymos		& {\it 1.1(2)}	& 2.9(4)		& 0.22(2)		& 11.91(1)	& [P03b][P05a]		\\ %
(85938) 1999 DJ$_4$		& {\it 0.7}		& {\it 3}		& 0.5(1)		& 17.73(1)	& [P04][Be04][P05a] 	\\ %
2005 AB				&			&			&{\it $\ge$0.24}&{\it 17.9}	& [Rd05]			\\ %
\\\hline
  \end{tabular}
 \end{center}
 Published uncertainties in least significant digit(s) shown in parentheses.  Estimated quantities or values published without error estimates shown in italics.  \\
 $^1$ Selected references shown.  For more complete references see Richardson \& Walsh (2005) and Pravec et al.\ (2005a).\\ 
 $^2$ Method resulting in discovery.  Many objects have been observed by both radar and lightcurve techniques. 
 
[Be01] Benner et al.\ 2001, [Be03] Benner et al.\ 2003, [Be04] Benner et al.\ 2004, 
[Mg02] Margot et al.\ 2002,  [Mg03]  Margot et al.\ 2003,  [Mo97] Mottola et al.\ 1997, [ML00] Mottola \& Lahulla 2000, [Ne03] Neish et al.\ 2003, [No00] Nolan et al.\ 2000, [No02a,b]  Nolan et al.\ 2002a,b, [No03] Nolan et al.\ 2003,  [No04]  Nolan et al.\ 2004,  [Os03] Ostro et al.\ 2003, 
[PH97] Pravec \& Hahn 1997, [P98] Pravec et al.\ 1998, [P00b] Pravec et al.\ 2000b,   [PS01] Pravec \& Sarounova 2001, [P02] Pravec et al.\ 2002, [P03] Pravec et al.\ 2003, [P04] Pravec et al.\ 2004, [P05a] Pravec et al.\ 2005a,
 [Sh04] Shepard et al.\ 2004,  [Sl04] Schlieder et al.\ 2004,
 [Rd05] Reddy et al.\ 2005, [WPP05] Warner, Pravec, \& Pray 2005
\end{table}
\begin{table}\def~{\hphantom{0}}
  \begin{center}
  \caption{Main Belt Binaries}
  \label{tab:mbb}
  \begin{tabular}{lllllr}\hline
      			  	& \multispan2 {semimajor~axis\ }	&			& 	Period	& 	reference$^1$		\\
 Object      		& a (km) 		& a/r$_p$  		&  r$_s$/r$_p$ 	& \ \ (days) \ \ 	& 					\\\hline
{\em imaging}		&			& 				& 			&			&					\\
(243) Ida			& {\it 108}	& {\it 7.0}			& {\it 0.045}	& {\it 1.54}	& [BC94][Me02]		\\ 
(45) Eugenia		& 1196(4)		& {\it 11.1}		& {\it 0.06}	& 4.7244(10)	& [Me99b][Me02][Ma04]	\\ 
(90) Antiope		& 170(1)		& 3.1(5)	 		& {\it 0.99}	& 0.68862(5)	& [Me00a]	[De05]		\\ 
(762) Pulcova		& {\it 810}	& {\it 11.6}		& {\it 0.14}	& {\it 4.0}		& [Me00b][Me02]		\\ 
(87) Sylvia$^2$		& 1356(5)		& 17.6(7)			& 0.12(2)		& 3.6496(7)	& [Br01][St01a][Ma05a]	\\
				& 706(5)		&  9.2(4)			& 0.045(15)	& 1.3788(7)	& [Ma05a] 			\\ 
(107) Camilla		& 1235(16)	& {\it 11}			& 0.040(4)	& 3.710(1)	& [St01b][Ma05b]		\\ 
(22) Kalliope		& 1065(8)		& 11.8(4)			& 0.22(2)		& 3.590(1)	& [Me01a],[MB01][Ma03] \\ 
(3749) Balam		& 310(20)		& {\it 52}			& {\it 0.22}	& 110(25)		& [Me02a]	[Ma05b]		\\ 
(121) Hermione		& 768(11)		& 7.4(3)			& {\it 0.06}	& 2.582(2)	& [Me02b][Ma05c]		\\ 
(1509) Escalonga	& 			& 				& {\it 0.33}	&			& [Me03a]				\\ 
(283) Emma		& 596(3)		& {\it 8}			& {\it 0.08}	& 3.360(1)	& [Me03b][Ma05b]		\\ 
(379) Huenna		& 3400(11)	& {\it 74} 			& {\it 0.08}	& 80.8(4)		& [Mg03b][Ma05b]		\\ 
(130) Elektra		& 1252(30)	& {\it 13.5}		& {\it 0.02}	& 3.92(3)		& [Me03c]	[Ma05b]		\\ 
(22899) 1999 TO$_{14}$& 		& 				& {\it 0.3}		& 			& [Me03d]			\\ 
(17246) 2000 GL$_{74}$& 		& 				& {\it 0.4}		&			& [Tm04]				\\ 
(4674) Pauling		& 			& 				& {\it 0.3}		&			& [Me04a]				\\ 
				&			& 				&	 		&			&					\\
{\em lightcurve}	&			& 				&	 		&			&					\\
(3782) Celle		& 			& 				& 0.42(2)		& 1.5238(13)	& [Ry03]				\\ 
(1089) Tama		& 			& 				& {\it 0.7}		& 0.6852(2)	& [Bh04a]				\\ 
(1313) Berna		& 			& 				& 			& 1.061(5)	& [Bh04b]				\\ 
(4492) Debussy		& 			&				& 			& {\it 1.108}	& [Bh04c]				\\ 
(854) Frostia		&			& 				& {\it 0.4}		& {\it 1.565}	& [Bh04d]				\\ 
(5905) Johnson		&			& 				& 0.40(4)		& 0.907708(2)	& [W05a,c]			\\ 
(76818) 2000 RG$_{79}$&		&				& 0.37(3)		& 0.5885(4)	& [W05b]				\\ 
(3982) Kastel		&			&				&			&			& [P05b]				\\ 
(809) Lundia		&			&				&{\it 1}		& {\it 0.64}	& [Kr05]				\\ 
(9069) Hovland		&			& 				&{\it 0.3}		&			& [W05c]				\\ 
				&			& 				&	 		&			&					\\
{\em Trojan}		&			& 				&	 		&			&					\\
(617) Patroclus)		& 685(40)		& 11(1)			& 0.92(5)	 	& 2.391(3) {\em or}	& [Me01b][Ma05d]	\\ 
				&			& 				&	 		& 4.287(2)		&				\\
\\\hline
  \end{tabular}
 \end{center}

Published uncertainties in least significant digit(s) shown in parentheses.  Estimated quantities or other values published without error estimates shown in italics.\\
 $^1$ Selected references shown.  For additional references see Merline et al.\ (2002) or Richardson \& Walsh (2005).\\
 $^2$ Average of long and short ellipsoid axes used for r.
 
 [BC94] Belton \& Carlson 1994, [Br01] Brown et al.\ 2001, [De05] Descamps et al.\ 2005, [Kr05] Kryszczynska et al.\ 2005,
 [Ma03] Marchis et al.\ 2003, [Ma04] Marchis et al.\ 2004, [Ma05a,b,c,d] Marchis et al.\ 2005a,b,c,d,
 [MB01] Margot \& Brown 2001, 
 [Me99] Merline et al.\ 1999, [Me00a,b] Merline et al.\ 2000a,b, [Me01a,b] Merline et al.\ 2001a,b, [Me02a,b] Merline et al.\ 2002a,b,  
 [Me03a,b,c,d] Merline et al.\ 2003a,b,c,d, [Me04a] Merline et al.\ 2004a,
 [P05b] Pravec et al.\ 2005b, 
[Ry03] Ryan et al.\ 2003, [St01a,b] Storrs et al.\ 2001a,b, [Tm04] Tamblyn et al.\ 2004,
[W05a,b,c] Warner et al.\ 2005a,b,c 
\end{table}
\begin{table}\def~{\hphantom{0}}
  \begin{center}
  \caption{Transneptunian Binaries}
  \label{tab:tnb}
  \begin{tabular}{llllllr}\hline
      								&\multispan2{semimajor~axis\ }	&			&			& 	Period		& reference$^1$ 	\\
 Object      						& a (km) 			& a/r$_p$ 	&  r$_s$/r$_p$	&  \ \ e \ \ 		& \ \ (days) \ \		&  				\\\hline
%
{\em classical }						&				&		&			&			&				&				\\
(88611) 2001 QT$_{297}$				& 27,300(340)		&{\it 410}	&{\it 0.72}	& 0.240(3)	& 825(3)			& [Op03] [K05]		\\
\phantom{(12345) }1998 WW$_{31}$	& 22,300(800)		&{\it 300}	&{\it 0.83}	& 0.82(5)		& 574(10)			& [V02] 			\\
(58534) 1997 CQ$_{29}$				& 8,010(80)		&{\it 200}	&{\it 0.91}	& 0.45(3)		& 312(3)			& [N04a] 			\\
(66652) 1999 RZ$_{253}$				& 4,660(170)		&{\it 56}	&{\it 1.0}		& 0.460(13)	& 46.263(6/74)		& [N04b]  		\\
\phantom{(12345) }2001 QW$_{322}$	&				&		&{\it 1.0}		&			&				& [Kv01] 			\\
\phantom{(12345) }2000 CF$_{105}$		&				&		&{\it 0.73}	&			& 				& [N02a][N02b]	\\
\phantom{(12345) }2000 CQ$_{114}$	&				&		&{\it 0.81}	&			&				& [SNG04] 		\\
\phantom{(12345) }2003 UN$_{284}$	&				&		&{\it 0.76}	&			&				& [MC03] 		\\
\phantom{(12345) }2003 QY$_{90}$		&				&		&{\it 0.95}	&			&				& [EKC03] 		\\
\phantom{(12345) }2005 EO$_{304}$	&				&		&{\it 0.58}	&			&				& [KE05] 			\\
\CM								&				&		&{\it 0.58}	&			&				& [SN05] 			\\
\phantom{(12345) }\OJS				&				&		&{\it 0.69}	&			&				& [SN05]			\\
\CS								&				&		&{\it 0.95}	&			&				& [SN05]			\\
\phantom{(12345) }\OJ				&				&		&{\it 0.54}	&			&				& [SN05]			\\
%
{\em scattered }						&				&		&			&			&				&				\\
\phantom{(12345) }2003 EL$_{61}$		& 49,500(400)		&		&{\it0.22}		& 0.050(3)	& 49.12(3)		& [Br05a,b] 			\\
\phantom{(12345) }2001 QC$_{298}$	& 3,690(70)		&		&{\it0.79}		& 			& 19.2(2)			& [Mg04] 		 	\\
\YW								&				&		&{\it0.55}		&			&				& [SN05] 			\\
\TL								&				&		&{\it0.46}		&			&				& [SN05]			\\
\phantom{(12345) }2003 UB$_{313}$     	&				&		&{\it0.13}		&			&				& [Br05c,d] 			\\
%
{\em resonant }						&				&		&			&			&				&				\\
\phantom{(12345)}Pluto/Charon			& 19,636(8)		&  16.7(4)	& 0.53(3)		& 0.0076(5)	& 6.38722(2)		& [TB97]	 		\\
(26308) 1998 SM$_{165}$				& 11,310(110)		&		&{\it0.42}		&			& 130(1)			& [BT02][Mg04]		\\
(47171) 1999 TC$_{36}$				& 7,640(460)		&		&{\it0.39}		&			& 50.4(5)			& [TB02][Ma02][Mg04]\\	

\\\hline
  \end{tabular}
 \end{center}
Published uncertainties in least significant digit(s) shown in parentheses.  Estimated quantities or other values published without error estimates shown in italics.

$^1$ Selected references shown. 

[BT02] Brown \& Trujillo 2002, [Br05a,b,c,d] Brown et al.\ 2005a,b,c,d,  [EKC03] Elliot, Kern \& Clancy 2003, [Kv01] Kavelaars et al.\ 2001, [KE05] Kern \& Elliot 2005, [K05] Kern 2005 thesis, [Ma02] Marchis et al. 2002, [Mg04] Margot et al.\ 2004, [MC03] Millis \& Clancy 2003, [N02a,b] Noll et al.\ 2002a,b, [N04a,b] Noll et al.\ 2004a,b,  [Op03] Osip et al.\ 2003, [SNG04] Stephens, Noll \& Grundy 2004, [SN05] Stephens \& Noll 2005, [TB97] Tholen \& Buie 1997, [V02] Veillet et al.\ 2002
  
\end{table}

\subsection{Population statistics}

The number of binaries is still quite small for any statistical studies, particularly studies related to the binary fraction in different subpopulations.  There are, however, interesting hints that are becoming apparent with the current inventory (Table~\ref{tab:freq}).  As history has shown with other studies of minor bodies, as the number of discoveries increase, important patterns can be expected to emerge.  

Any discussion of the frequency of binaries must take into account observational limits and biases.  This is particularly important in the Kuiper Belt where the separations of the majority of binaries are not resolved by ground-based observations and among the smaller radii families in the Main Belt where potential binaries are extremely faint.  

Gross population statistics are most frequently cited for each of the major small-body populations, although, as discussed below this is probably an oversimplification.  In the most recent and extensive work on the subject, Pravec et al.\ (2005a) find 15$\pm4$\% of NEAs are binary (11 binaries) with their lightcurve survey sensitive to primaries with $r_p > 0.15 $km and $r_s/r_p > 0.18$.  Margot et al.\ (2002) found 5 binary companions of primaries larger than 0.2 km in diameter in their sample of ``$\sim$50" NEAs studied by radar.  The authors quote a fraction of near 16\% which implies that 31 of the 50 meet the diameter criterion.  Though not stated in Margot et al., the small size of the sample implies a relatively large uncertainty of +9/-5\%.  Of the remaining 19 smaller primaries, none were found to have binaries which I use to estimate an upper limit of $< 9$\%, although this is likely subject to observational bias.   Interestingly, the frequency of detected NEA binaries agrees with the number estimated from terrestrial double craters (Bottke \& Melosh 1996).  Despite this apparent agreement, however, there is clearly room for improvement in the statistics and for studies of the fraction of NEA binaries as a function of size, spin, and orbit characteristics.  It will be especially interesting to be able to compare binary frequency in NEAs with the frequency in comparably sized Main Belt populations.

In the Main Belt, Merline et al.\ (2002) report $\sim$2\% of the 300 objects in their Main Belt survey have relatively large (``tens of km") binaries.   Of the 6 Trojans searched by Merline et al.\ (2002), one, Patroclus, was found to be binary.  The sample size is insufficient to determine the binary frequency, but ongoing search programs (e.g. Merline et al., HST cycle 14 program 10512) may remedy this shortcoming.

A bewildering variety of frequencies are cited for the transneptunian population, some based on gross statistics of the number of known binaries divided by the number of known TNOs.  However, this is a particularly inaccurate method for estimating the frequency of TNO binaries since many are undetectable by typical ground-based observations.  Of the 22 known TNBs, 13 have been found with the Hubble Space Telescope.  The other 8 have been found at ground based telescopes all with separations greater than 0.3 arcsec.  The fraction of objects with widely separated, relatively bright companions is clearly small.  Schaller \& Brown (2003) failed to find any companions in a sample of 150 TNOs observed with Keck.  This fact can also be deduced from the relatively small fraction of objects found by groundbased telescopes, 8, out of the more than 1000 separate TNOs so far observed at least once.  However, the inhomogeneity of these ground based observations precludes any quantitative analysis of these statistics.  A subset of the HST-discovered objects come from a uniform sample observed with the NIC2 camera and can provide a better estimate of the global frequency of binaries of 11\% $\pm {5\atop 2}$ (Stephens \& Noll 2005). 

Neither Kuiper Belt, Main Belt, nor NEAs are homogeneous populations.  It is, therefore, naive to expect that the frequency of binaries in each of the distinct subpopulations will be identical.  If they are not, then the gross statistics in the previous paragraphs have marginal utility because they depend on the makeup of the sample.  In the case of transneptunians, the populations are identified by their orbital dynamics.  Distinct populations identified are the classical Kuiper belt (nonresonant, cubewanos), the resonant objects (including, but not limited to the Plutinos in the 3:2 resonance), and the scattered disk objects.  The classical belt is further divided into two overlapping populations, one, the cold classical disk, have inclinations relative to the Kuiper Belt plane of less than 5 degrees.  The hot classicals are identified by their inclinations in excess of this cutoff.  The scattered disk consists of both near and extended scattered groups.   In the Main Belt, the subpopulations of most interest are collisional families, but there may also be differences in binary frequency with taxonomical class and size.  For the NEAs there may also be differences as a function of size and orbital dynamics.

An analysis of 84 TNOs observed with NICMOS has revealed a total of 9 previously unknown binaries (Stephens \& Noll 2005).   This sample is large enough that it is possible, for the first time, to identify a factor of four higher fraction of binaries in the cold classical disk than in the combined dynamically excited populations.    The statistics for other dynamical classes of TNOs remains too small for further discrimination, but ongoing observations with HST may result in rapid progress.
Reporting on the detection of the companion to 2003 UB$_{313}$, Brown et al.\ (2005d) note that 3 of 4 of the brightest TNOs (Pluto, 2005 FY$_9$, 2003 EL$_{61}$, and 2003 UB$_{313}$) have satellites.  However, because satellites 3.2-4.2 magnitudes fainter than their primary (companions of 2003 EL$_{61}$, and 2003 UB$_{313}$ respectively) would not have been detected in the large NICMOS survey (Stephens \& Noll 2005) it is premature to conclude that large TNOs may have a higher fraction of binaries.  Deeper observations of a substantial number of smaller, fainter, TNOs will be needed to test the hypothesis of a size-dependence in binary frequency.

Several studies have now begun to focus on the search for binaries within specific collisional families within the Main Belt.  Merline et al.\ (2004b) reported the discovery of 2 Koronis binaries out of a sample of 9 while finding no binaries among 17 Karin cluster targets and 18 Veritas family targets, all observed with HST.  This is a significantly larger fraction than would be expected in a population with only 2\% binaries, though it remains subject to small number uncertainties.  Merline et al.\ (2004b) speculated that the fraction of binaries might depend on the size of the parent body and the ability of the subsequent ejecta to form bound pairs.  An alternative explanation is that the binary frequency in a collisional family increases as the size relative to the largest fragment decreases.  In this particular case, the searches in the Koronis family are sampling relatively smaller members of the family.  The apparent higher proportion of binaries in the Koronis family appears to contradict another hypothesis that newer collisional families would have more binaries than old families.  Colas et al.\ (2005) report the detection of 4 binaries among a sample of $\sim$40 asteroids with diameters between 10 and 50 km.  This is suggestive of a difference in binary frequency as a function of size, but, as with the other families, the statistics are still weak.

\begin{table}
  \begin{center}
  \caption{Fraction of Binaries in Minor Planet Populations}
  \label{tab:freq}
  \begin{tabular}{lll}\hline
  Population   				& binaries (\%)          		& reference \\\hline
\em{NEA}				&						&						\\
$d > 0.3$ km				& 15$\pm{4}$				& Pravec et al.\ 2005a			\\
$d > 0.3$ km, $2.2 < $P$ < 2.8$ hr & 66$\pm{10\atop12}$		& \ \ \ \ \ "					\\
						&						&						\\
$d > 0.2$ km				& 16$\pm{9\atop5}^*$		&  Margot et al.\ 2002		\\
$d < 0.2$ km				& $< 9^*$					&  \ \ \ \ \ 	"				\\
						&						&						\\
\em{Main Belt}				&						&						\\
``Average"					& $\sim$2					&  Merline et al.\ 2002		\\
 10km $< d <$ 50km			& $\sim10\pm{7\atop3}^*$	&  Colas et al.\ 2005			\\
Koronis 					& 22$\pm{18\atop9}^*$		&  Merline et al.\ 2004b		\\
Karin					& $< 10^*$				& \ \ \ \ \	"				\\
Veritas					& $< 9^*$					& \ \ \ \ \ 	"				\\
Vestoids					&						& Ryan et al.\ 2004			\\
Hungarias 					&						& Warner et al.\ 2005			\\
						&						&						\\
\em{Transneptunian}			&						&						\\
 ``Average" 				& 11$\pm{5\atop2}$ 			&  Stephens \& Noll 2005 		\\
 Cold Classical 				& 22$\pm{10\atop5}$	 	& \ \ \ \ \    "		 		\\
 All other 					& 5.5$\pm{4\atop2}$ 		&  \ \ \ \ \   "                            	\\
                  				&						&						\\
``Bright" TNOs				& 75$\pm{10\atop30}^*$		& Brown et al.\ 2005d		\\		 
 \\\hline
\end{tabular} 
\end{center}
$^*$ Uncertainties not reported, calculated in this work based on available information.
\end{table}

\section{Physical Properties: mass, albedo, and density}

One of the most practical benefits of binaries is the ability to derive the system mass from observation of the orbit.  With observations at multiple epochs and the centuries-old Kepler's laws, the determination of mass is conceptually simple.  In practice, solutions are found through iterative numerical methods.  Because of the difficulties in acquiring large blocks of observing time on major facilities, in many cases data are very sparse.  Specialized methods for constraining orbit solutions with limited data have been developed (Herstroffer \& Vachier\ 2005).   Orbits, and hence masses, have been obtained for only a subset of the known binaries.  For systems found through lightcurve analysis, the period and relative sizes of the components are determined, but the semimajor axis, $a$, of the mutual orbit is not.  Lacking $a$, the mass cannot be directly determined, though in some cases with modelling, reasonable guesses can be made (e.g.~Pravec et al.\ 2005a).  The values listed in Table~\ref{tab:phys} are limited to those objects with directly measured orbits.

An important question is whether brightness can be used as a proxy for mass.  This depends on the uniformity of the albedo of the objects in the minor body population in question.  These are not known a priori.  For NEAs and Main Belt asteroids, it is usually possible to directly measure the diameter of the objects.  In that case, it is a relatively simple step to determine an albedo from the measured apparent magnitude and known distance.  Complications arise, particularly for NEAs and smaller Main Belt asteroids due to the effects of non-spherical shape, phase function, and complex viewing geometries.  Spectral taxonomy can used to estimate albedo for objects with unknown diameters.  Transneptunian objects are too distant to be resolved.  Measurement of thermally emitted flux coupled to observed reflected flux and models of rotation and thermal inertia can be used to determine albedo.  Only a few TNOs have had their albedos measured in this way from the ground and from the Spitzer Space Telescope (Grundy et al.\ 2005a,b).  Interestingly, the determination of a mass for binaries allows one to constrain the albedo as a function of assumed density.  This approach can provide significant constraints on albedo because it varies as the 2/3 power of assumed density;  over a ``reasonable" range of density (500 - 2000 kg/m$^3$) albedos vary by a factor of 2.5 (Noll et al.\ 2004a).  So far, there are no apparent correlations of TNO albedo with any other observable (Grundy et al.\ 2005a), so estimates by proxy are currently not possible.

The determination of density has the same limitations as does the determination of albedos.  If direct size determinations are available, density is calculable once the mass is known.  Sizes are best determined for NEAs and Main Belt asteroids since these can be resolved by radar or direct imaging.  Non-spherical shape dominates the uncertainties for NEAs to the degree that measured densities are frequently so uncertain as to be unconstraining (Table~\ref{tab:phys}).  Main Belt binaries result in the best density constraints because good diameters can be measured for the primary and because many of the systems are very asymmetric with most of the mass in the primary so that assumptions about the relative albedo of the primary and secondary introduce negligible uncertainty.  For both the Main Belt and the NEA populations the densities that have been determined so far are consistent with expected densities based on meteorite samples and reasonable bulk porosities (references).   The lack of measured diameters impedes the determination of densities in the transneptunian population.  Aside from the well-constrained Pluto/Charon binary, only one object has an estimated density, \TC\ (Stansberry et al.\ 2005).  Interestingly, the density of \TC\ appears to be extremely low, $\rho =$ 500-900 kg/m$^3$ implying a significant mismatch between surface area and filled volume.

\begin{table}\def~{\hphantom{0}}
  \begin{center}
  \caption{Mass and Density of Solar System Binaries}
  \label{tab:phys}
  \begin{tabular}{llll}\hline
 Object      					& system mass	 	 		&  density			& references 		\\
							&						& (g/cm$^3$)		&				\\\hline
{\it NEAs}					&($10^{9}$kg = Gg)			&				&				\\
(66391) 1999 KW$_4$			& 2330(230)				& 2.6(1.6)			& [Me02]			\\ %
2000 DP$_{107}$				& 460(50)					& 1.7(1.1)			& [Me02]			\\ %
2000 UG$_{11}$				& 5.1(5)					& 0.8(0.6)			& [Me02]			\\ %
(5381) Sekhmet    				& 						& 2.0(0.7)			& [Ne03]			\\ %
(3671) Dionysus					&						& {\it 1.0-1.6(3/2)}	& [P05a]			\\ %
(65803) Didymos				&						& {\it 1.7-2.1(3/2)}	& [P05a]			\\ %
1991 VH						&						& {\it 1.4-1.6(5)}	& [P05a]			\\ %
1996 FG$_3$					&						& {\it 1.3(6)}		& [P05a]			\\ %
							&						&				&				\\
{\it Main Belt}					& ($10^{18}$kg = Zg)		&				&				\\
(87) Sylvia					& 14.78(6) 				& 1.2(1)			& [Ma05a]			\\ %
(107) Camilla					& 10.8(4)$^1$				& {\it 1.88}		& [Ma05b]		\\ %
(130) Elektra					& 10.1(9)$^1$				& 3.8(3)			& [Ma05b]		\\ %
(22) Kalliope					& 7.3(2)$^1$				& 2.03(16)		& [Ma03]			\\ %
(45) Eugenia					& 5.8(2)$^1$				& 1.1(3)			& [Ma05b]		\\ %
(121) Hermione					& 5.4(2)$^1$				& 1.1(3)			& [Ma05c]			\\ %
762 Pulcova					& {\it 2.6}					& 1.8(8)			& [Me02]			\\ %
(283) Emma					& 1.49(2)					& 0.87(2)			& [Ma05b]		\\ %
(90) Antiope					& 0.82(1)					& 0.6(2)			& [De05]			\\ %
(379) Huenna					& 0.477(5)				& 1.16(13)		& [Ma05b]		\\ %
243 Ida						& {\it 0.042}				& 2.6(5)			& [Me02]			\\ %
							&						&				&				\\
{\it Trojan}					& ($10^{18}$kg = Zg)		&				&				\\
617 Patroclus					& {\it 1.5} 				& 1.3(5)			& [Me02]			\\ %
							& 1.4(2){\em or}			& 0.8(1){\em or} 	& [Ma05d]		\\ %
							& 4.3(2)					& 2.6(1)			& \ \ \	"		\\ %
							&						&				&				\\
{\it Transneptunian}				&($10^{18}$ kg = Zg)		&				&				\\
Pluto/Charon					& 14,710(20)				& 1.99(7)/1.66(15)	& [TB97]			\\ %
\phantom{(12345) }2003 EL$_{61}$	& 4,200(100)				&				& [Br05b]			\\ %
(47171) 1999 TC$_{36}$			&13.9(2.5)				& {\it 0.5-0.9} 		& [Mg04][Mg05][Sb05]\\ %
\phantom{(12345) }2001 QC$_{298}$&10.8(7)					&				& [Mg04]			\\ %
(26308) 1998 SM$_{165}$			& 6.78(24)				&				& [Mg04]			\\ %
(66652) 1999 RZ$_{253}$			& 3.7(4)					&				& [N04b]			\\ %
\phantom{(12345) }1998 WW$_{31}$& 2.7(4)					&				& [V02]			\\ %
(88611) 2001 QT$_{297}$	          	& 2.3(1)					&				& [Op03][K05]		\\ %
(58534) 1997 CQ$_{29}$			& 0.42(2)					&				& [N04a]			\\ %

\\\hline
  \end{tabular}
 \end{center}
 $^1$  Mass calculated from measured period and semimajor axis.
 
[Br05b] Brown et al.\ 2005, [De05] Descamps et al.\ 2005, [K05] Kern 2005, [Ma03] Marchis et al.\ 2003, [Ma04] Marchis et al.\ 2004, [Ma05a,b,c,d] Marchis et al.\ 2005a,b,c,d, [Mg04] Margot et al.\ 2004, [Mg05] Margot et al.\ 2005, [Me02] Merline et al.\ 2002, [Ne03] Neish et al.\ 2003, [N04a,b] Noll et al.\ 2004a,b,
[P05a] Pravec et al.\ 2005a, [Sb05] Stansberry et al.\ 2005, [Op03] Osip et al.\ 2003,  [V02] Veillet et al.\ 2002 
\end{table}

\begin{figure}
 \includegraphics{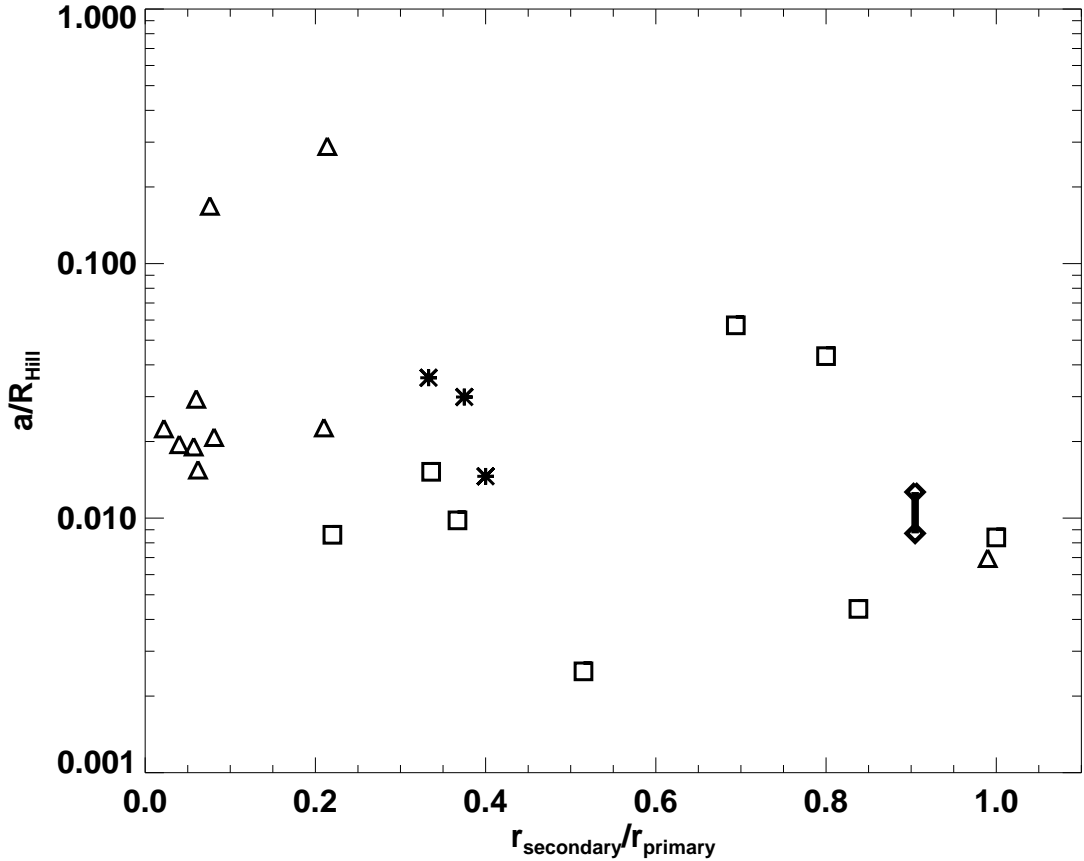}
  \caption{The characteristics of binaries in different solar system populations are shown in this figure.  (asterisks=NEAs, triangles=MainBelt, diamond=Trojan, squares=TNOs)  The separation, in terms of the fraction of the Hill radius is plotted against the size ratio of the binary pair.  We have plotted only objects for which a system mass determination has been made from observed orbital data (Tables~\ref{tab:neb}-\ref{tab:tnb},\ref{tab:phys}).  All of the binary populations have most objects within a similar small fraction of the Hill radius.  The main difference between populations is the large number of objects in the Kuiper Belt with large size ratios.  The lack of TNBs with small secondary to primary size ratios is heavily influenced by observing bias and the number of such systems remains to be determined.  It is clear that large primary to secondary ratio systems are rare in the Main Belt and among NEAs.}
    \hbox{} \label{fig:hill}
\end{figure}

\section{Formation and Destruction of Binaries}

An important area of investigation is the effort to understand when, how, and under what conditions binaries and multiples are formed and how long they are able to survive.  Theoretical work to date has identified three main formation mechanisms: capture, collisions, and rotational fission.  Currently, the leading models for binary formation differ for each of the major populations.  Richardson \& Walsh (2005) have thoroughly reviewed this topic and I include here only a few of the most salient points.

TNBs have had the greatest variety of models proposed for their formation.  Stern (2002) proposed a collisional model but noted that the then assumed fraction of 1\% binaries could not be met unless he assumed that the mean albedo was higher than the formerly assumed value of $p=0.04$ adopted from comet Halley.  While it is now apparent that albedos of many TNOs are indeed higher than 4\%, the collisional model cannot produce the much higher fraction of binaries that are now being found.  A collision model has also been proposed for the Pluto/Charon binary (Canup\ 2005) but stochastic models such as this have little applicability to larger populations of binaries.  Capture models have evolved in complexity with the most recent work by Astakhov et al.\ (2005) using chaos-assisted capture giving a reasonable distribution of orbital parameters compared to the observed sample.  The hybrid collision-exchange capture model of Funato et al.\ (2004) can now be ruled out because the eccentricity of observed orbits does not meet the predictions of this model.  

In the Main Belt, collisions appear to be the dominant mode of forming binaries.  The evidence for this is the small secondary to primary size ratio for most of the objects (Table~\ref{tab:mbb}) and the apparent higher incidence of binaries in collision families (Merline et al.\ 2004b).  Numerical models of formation of binaries (e.g.~Durda et al.\ 2004) identify two main classes of post-collision binaries: fragments that are captured or reaccreted around a remnant primary and pairs (or more) of fragments that are mutually captured.   

NEA binaries are distinctive in the rapid rotation of the primaries, typically less than 3 hours (Pravec et al.\ 2005a).  This is taken to be an indication that these objects may have formed through spin-up (from the YORP effect or collisions) and subsequent fission.  The role of tidal forces during close encounters with Earth and Mars has usually been assumed (Richardson et al.\ 1998, Walsh \& Richardson 2005).  However, the recent evidence for a similarly large fraction of binaries among rapidly-rotating small asteroids populations in the Main Belt opens the question of whether NEA binaries are formed in the Main Belt and survive the orbital perturbations that bring them into NEA orbits or whether binaries are efficiently formed during the relatively brief 10\ Myr mean lifetime of NEAs.  

Survival of binaries, once they are formed, is an important, and so far largely-neglected question.  Petit \& Mousis (2004) have investigated the survival of TNBs and have found that some of the most widely separated may have lifetimes shorter than the age of the solar system.  If this is correct, the initial inventory of binaries may have been even higher than the already high fraction we see today.  An important question, in addition to lifetime against collisional erosion, is the survival of binaries during scattering events.  The fact that most binaries orbit within 10\% of the Hill radius or less (Figure~\ref{fig:hill}), regardless of population, may be the signature of tidal disruption of the most loosely bound systems.

Given the large size range and potentially different histories of these populations, the variety of formation mechanisms may not entirely surprising, but from an aesthetic point of view, it is clearly less than satisfactory.  A question for the future is whether or not there are underlying and unifying modes for formation that apply across all the classes of minor bodies in the solar system.  At present, collisions appear to be the most universal component of any such model, even if, in some cases, they are only part of the formation scenario.  The apparent prevalence of binaries in the Kuiper Belt raises the possibility that bound systems are commonplace in accreting dust disks and that further evolution is dominated by the destruction (or lack thereof) of preexisting binaries.  The most promising case for consolidation is for NEA binaries which may turn out to be survivors of a preexisting binary population in the Main Belt.

\section{Binaries Beyond}

The Main Belt asteroid (87) Sylvia has now been confirmed as the first known minor planet triple (Marchis et al.\ 2005a,b).  The discovery of triples was not entirely unexpected as numerical simulations of collisions produce triples and higher multiples as well as binaries, at least in the early post-collision time scale (Leinhardt \& Richardson 2005).   Stability of multiples is an important issue that will have to be addressed, but as demonstrated by the existence of stellar triples and higher-order multiples, stable configurations can be found.  Ultimately, the incidence of triples relative to binaries may help constrain models of formation, especially collisional models.

No multiples have been identified among NEAs or TNOs as of mid-October 2005.  However, \TC\ has been proposed as a potential triple as a way of understanding the apparent low density derived from Spitzer observations (Margot et al.\ 2005, Stansberry et al.\ 2005).  Interestingly, an unexplained large change in brightness on one night has been reported for this object (Ivanova et al.\ 2005).  Further monitoring of the lightcurve is called for.  The TNO \CQ\ also has reported large variations in one of the two components (Noll et al.\ 2004, Margot et al.\ 2005).  This may be the signature of a close binary or contact system, or may simply represent large albedo and shape effects.  The largest TNOs have larger Hill radii (for objects of the same density r$_H$ scales with the radius of the primary) and a larger fraction of their Hill spheres are searchable with instruments of a given resolution.  Stern (2003) reviewed searches for additional satellites of Pluto and  prospects for improved searches using the Hubble Space Telescope and the New Horizons spacecraft.  With a high fraction of Pluto-sized objects now known to have one satellite, it seems more likely that multiples will eventually be found.

The discovery of numerous binary systems has stimulated creative and unconventional thinking.  {\'C}uk \& Burns (2005) propose a variant of the YORP effect which would apply to binaries.  For NEAs, which are small enough and close enough to the Sun to be affected, this effect can explain the nearly circular and synchronous orbits of secondaries.  The timescale for orbital evolution under the BYORP effect is surprisingly short, generally less than 10$^5$ years.  Because of the short evolutional time scale, {\'C}uk and Burns speculate that most NEA binaries may be in stable states where YORP and BYORP cancel and are small.  This prediction is testable with accurate observations of orbital elements for NEA binary pairs, a test that should be possible within the next five years.  On a much grander scale, Cintala et al.\ (2005) suggest an "experiment" aimed at deorbiting a binary companion with multiple explosive charges.  They list candidates in order of $\Delta v$ needed to deorbit the secondary and crash it into the primary.  The authors identified 2000~UG$_{11}$ as the best candidate with a 230 m diameter primary and a 140 m secondary orbiting with a semimajor axis of 410 m.

Comet nuclei are thought to originate from the same pool of primordial bodies that populate the outer solar system and are thus likely to share a similar propensity for binaries.  Study of comet nuclei is a notoriously difficult problem because of the small size of the objects and because of the persistent presence of coma, even at large heliocentric distances.  Disruption of comet nuclei is observed, with perhaps the best known example being comet Shoemaker-Levy 9 (c.f. Weaver et al.\ 1995).  There are, however, no conclusive observations of a gravitationally bound cometary binary.  Marchis et al.\ (1999) speculated that some aspects of comet Hale-Bopp might be more easily explained if that object were a bound binary, but this was not a unique interpretation.  It seems likely, however, that some fraction of comets, particularly ``new" Oort-cloud comets could be binary systems.  Observations of nuclei at large heliocentric distances with high-angular resolution could address this question.

The immediate future of the study of binaries in minor planets is very promising.  Discovery will continue, probably at an increased pace, due both to the heightened awareness of the existence and detectability of these bodies and to the availability of instrumentation capable of finding faint companions at small separations.  Binaries will continue to provide critical physical data that is not obtainable in other way.  The growing number of detected binaries will enable the nascent study of binaries fraction as a function of dynamical history, composition, size, rotation rate, and other conceivable correlations.  There are many possible avenues of investigation.  Among the most promising are apparently high fraction of binaries in the dynamically unperturbed classical Kuiper Belt (Stephens \& Noll 2005) and the apparently high fraction of binaries in some Main Belt collisional families (Merline et al.\ 2005) because both have the potential to yield clues on the origin and survival of binaries. 

We have now reached the point where it is possible to state that binaries must be a common feature in the evolution of protoplanetary disks like the one that formed our solar system.  Debris disks that we now observe around other stars probably share an affinity for forming pairs.  It remains to be seen if the great utility of binaries as mass-measuring tools and in teasing out the early history of the solar system will result in insights that extend beyond to inform our understanding of planetary systems in general.  If nothing else, however, the triumph of the detection and utilization of minor planet binaries must stand as an object lesson in the value of persistence and as an especially poignant reminder of the unexpected complexity of planetary systems.

\section{Online Resources}

In a rapidly developing field such as this, online compilations are invaluable resources for researchers and students.  While online resources are typically not referenceable, because of their ephemeral nature, it is worth mentioning several resources that are currently available.  A comprehensive compilation of solar system binaries is regularly updated by W.~R.~Johnston at his website http://www.johnstonsarchive.net/astro/asteroid\-moons.html.  The Ondrejov Asteroid Photometry Project led by Petr Pravec maintains an online resource at http://sunkl.asu.cas.cz/$\sim$ppravec/ that includes prepublication data and an updated list of binary NEAs.  Joel Parker maintains the Distant EKOs (Edgeworth-Kuiper Belt Objects) web pages that include a compilation of transneptunian binaries.  Franck Marchis has an extensive online collection of binary asteroid orbital information, much of which has not been published at the time of this review.  Updated information is available at http://astron.berkeley.edu/$\sim$fmarchis/  and was used in the tables in this review; references, however, were to published materials.

\begin{acknowledgments}
I would like to acknowledge ongoing and productive collaborations with my colleagues, particularly W.~Grundy, D.~Stephens, D.~Osip, J.~Spencer, and M.~Buie.  A tip of the hat also to A.~Storrs and B.~Zellner, early collaborators who helped me appreciate the excitement of objects smaller than Saturn.  A special note of thanks goes to A.~Lubenow who was the key enabler for complex solar system programs on the Hubble Space Telescope for two decades and who did so with his unique mixture of deep technical expertise and intolerance for nonsense.  This work was supported in part by grant GO 10514 from the Space Telescope Science Institute which is operated by AURA under contract from NASA.
\end{acknowledgments}


\begin{thebibliography}{}

\bibitem[Andre(1901)]{1901AstronomischeNachtrichen} Andr\'e, C.~L.~F.\ 1901, Astronomische Nachtrichen, 155, 27

\bibitem[Astakhov et al.(2005)]{2005MNRAS.360..401A} Astakhov, S.~A., Lee, 
E.~A., \& Farrelly, D.\ 2005, MNRAS, 360, 401 

\bibitem[Behrend et al.(2004a)]{Bh04a} Behrend, R., et al.\ 
2004a, IAUC, 8265  

\bibitem[Behrend et al.(2004b)]{Bh04b} Behrend, R., et al.\ 
2004b, IAUC, 8292   

\bibitem[Behrend(2004c)]{Bh04c} Behrend, R.\ 2004c, IAUC, 
8354  

\bibitem[Behrend et al.(2004d)]{Bh04d} Behrend, R., 
Bernasconi, L., Klotz, A., \& Durkee, R.\ 2004, IAUC, 8389 

\bibitem[Belton \& Carlson(1994)]{BC94} Belton, M., \& 
Carlson, R.\ 1994,IAUC, 5948, 2 

\bibitem[Benner et al.(2001)]{Be01} Benner, L.~A.~M., Ostro, 
S.~J., Giorgini, J.~D., Jurgens, R.~F., Margot, J.~L., \& Nolan, M.~C.\ 
2001, IAUC, 7632 

\bibitem[Benner et al.(2003)]{Be03} Benner, L.~A.~M., Nolan, 
M.~C., Margot, J.~L., Ostro, S.~J., \& Giorgini, J.~D.\ 2003, AAS/Division 
for Planetary Sciences Meeting Abstracts, 35,  24.01

\bibitem[Benner et al.(2004)]{Be04} Benner, L.~A.~M., Nolan, 
M.~C., Ostro, S.~J., Giorgini, J.~D., Margot, J.~L., \& Magri, C.\ 2004, 
IAUC, 8329 
 
\bibitem[Bottke \& Melosh(1996)]{1996Icar..124..372B} Bottke, W.~F., \& 
Melosh, H.~J.\ 1996, Icarus, 124, 372 

\bibitem[Brown et al.(2001)]{Br01} Brown, M.~E., Margot, 
J.~L., de Pater, I., \& Roe, H.\ 2001, IAUC, 7588 

\bibitem[Brown et al.(2005a)]{Br05a} Brown, M.~E., Trujillo, 
C.~A., \& Rabinowitz, D.\ 2005a, IAUC, 8577 

\bibitem[Brown et al.(2005b)]{Br05b} Brown, M.~E., et al.\ 
2005b, ApJ, 632, L45 

\bibitem[Brown(2005c)]{Br05c} Brown, M.~E.\ 2005c, IAUC, 
8610 

\bibitem[Brown et al.(2005d)]{Br05d} Brown, M.~E., et al. \ 2005d, 
ApJ, submitted 

\bibitem[Canup(2005)]{2005Sci...307..546C} Canup, R.~M.\ 2005, Science, 
307, 546 

\bibitem[Christy \& Harrington(1978)]{1978AJ.....83.1005C} Christy, J.~W., 
\& Harrington, R.~S.\ 1978, AJ, 83, 1005 

\bibitem[Cintala et al.(2005)]{2005LPI....36.2160C} Cintala, M.~J., Durda, 
D.~D., \& Housen, K.~R.\ 2005, 36th Annual Lunar and Planetary Science 
Conference, 36, 2160 

\bibitem[Colas et al.(2005]{De05ACM}   Colas, F., Behrend, R., Klotz, A., 
Roy, R., \& Bernasconi, L.\ IAU Symp. 229, abstract

\bibitem[{\'C}uk \& Burns(2005)]{2005Icar..176..418C} {\'C}uk, M., \& 
Burns, J.~A.\ 2005, Icarus, 176, 418 

\bibitem[Descamps et al.(2005]{De05ACM}   Descamps, P., Marchis, F., 
Michalowski, T., Berthier, J., Hestroffer, D., Vachier, F., Colas, F., 
Birlan, M.\ IAU Symp. 229, abstract  

\bibitem[Durda et al.(2004)]{2004Icar..170..243D} Durda, D.~D., Bottke, 
W.~F., Enke, B.~L., Merline, W.~J., Asphaug, E., Richardson, D.~C., \& 
Leinhardt, Z.~M.\ 2004, Icarus, 170, 243 

\bibitem[Elliot et al.(2003)]{2003IAUC.8235....2E} Elliot, J.~L., Kern, 
S.~D., \& Clancy, K.~B.\ 2003, IAUC, 8235

\bibitem[Durda et al.(2004)]{2004Icar..170..243D} Durda, D.~D., Bottke, 
W.~F., Enke, B.~L., Merline, W.~J., Asphaug, E., Richardson, D.~C., \& 
Leinhardt, Z.~M.\ 2004, Icarus, 170, 243 
 
\bibitem[Funato et al.(2004)]{2004Natur.427..518F} Funato, Y., Makino, J., 
Hut, P., Kokubo, E., \& Kinoshita, D.\ 2004, Nature, 427, 518 
 
\bibitem[Goldreich et al.(2002)]{2002Natur.420..643G} Goldreich, P., 
Lithwick, Y., \& Sari, R.\ 2002, Nature, 420, 643 

\bibitem[Grundy et al.(2005a)]{2005Icar..176..184G} Grundy, W.~M., Noll, 
K.~S., \& Stephens, D.~C.\ 2005a, Icarus, 176, 184 

\bibitem[Grundy et al.(2005b)]{2005DPS....37.5207G} Grundy, W.~M., Spencer, 
J.~R., Stansberry, J.~A., Buie, M.~W., Chiang, E.~I., Cruikshank, D.~P., 
Millis, R.~L., \& Wasserman, L.~H.\ 2005b, AAS/Division for Planetary 
Sciences Meeting Abstracts, 37,  52.07

\bibitem[Herstroffer \& Vachier(2005]{De05ACM}   Herstroffer, D. \& 
Vachier, F.\ IAU Symp. 229, abstract 

\bibitem[Ivanova et al.(2005)]{ACMabstractP6.11} Ivanova, V., Borisov, G., 
\& Belskaya, I.\ 2005, IAU Symp. 229, abstract 

\bibitem[Kavelaars et al.(2001)]{2001IAUC.7749....1K} Kavelaars, J.~J., 
Petit, J.-M., Gladman, B., \& Holman, M.\ 2001, IAUC, 7749 

\bibitem[Kern \& Elliot(2005)]{2005IAUC.8526....2K} Kern, S.~D., \& Elliot, 
J.~L.\ 2005, IAUC, 8526

\bibitem[Kern (2005)]{K05} Kern, S.~D.\ 2005, PhD thesis, MIT

\bibitem[Kryszczynska et al.(2005)]{Kr05} Kryszczynska, A., Kwiatkowski, 
T., Hirsch, R., Polinska, M., Kaminski, K. \&  Marciniak, A.\ 2005, 
CBET, 239  

\bibitem[Leinhardt \& Richardson(2005)]{2005Icar..176..432L} Leinhardt, 
Z.~M., \& Richardson, D.~C.\ 2005, Icarus, 176, 432 

\bibitem[Marchis et al.(1999)]{1999A&A...349..985M} Marchis, F., 
Boehnhardt, H., Hainaut, O.~R., \& Le Mignant, D.\ 1999, A\&Ap, 349, 985 

\bibitem[Marchis et al.(2003)]{Ma03} Marchis, F., Descamps, 
P., Hestroffer, D., Berthier, J., Vachier, F., Boccaletti, A., de Pater, 
I., \& Gavel, D.\ 2003, Icarus, 165, 112 

\bibitem[Marchis et al.(2004)]{Ma04} Marchis, F., Descamps, 
P., Hestroffer, D., Berthier, J., \& de Pater, I.\ 2004, AAS/Division for 
Planetary Sciences Meeting Abstracts, 36,  46.02 

\bibitem[Marchis et al.(2005a)]{Ma05a} Marchis, F., Descamps, 
P., Hestroffer, D., \& Berthier, J.\ 2005a, Nature, 436, 822  

\bibitem[Marchis et al.(2005b)]{Ma05b}  Marchis , F., Berthier, J.,  Clergeon, C.,  
Descamps, P.,  Hestroffer, D., de Pater, I., Vachier, F.\ 2005b, IAU 
Symp. 229, abstract  

\bibitem[Marchis et al.(2005c)]{Icarusinpress}  Marchis, F.,  Hestroffer, D., 
Descamps, P., Berthier, J., Laver, C., \& de Pater, I.\ 2005c, Icarus, in press 
   
\bibitem[Marchis et al.(2005d)]{2005DPS....37.1407M} Marchis, F., et al.\ 
2005d, AAS/Division for Planetary Sciences Meeting Abstracts, 37,  14.07 


\bibitem[Margot \& Brown(2001)]{MB01} Margot, J.-L., \& Brown, 
M.~E.\ 2001, IAUC, 7703 

\bibitem[Margot et al.(2002)]{Mg02} Margot, J.~L., Nolan, 
M.~C., Benner, L.~A.~M., Ostro, S.~J., Jurgens, R.~F., Giorgini, J.~D., 
Slade, M.~A., \& Campbell, D.~B.\ 2002, Science, 296, 1445 

\bibitem[Margot \& Keck(2003b)]{Mg03b} Margot, J.~L., \& Keck, 
W.~M.\ 2003b, IAUC, 8182

\bibitem[Margot et al.(2003a)]{Mg03a} Margot, J.~L., et al.\ 
2003a, IAUC, 8227 
 
\bibitem[Margot et al.(2004)]{2004DPS....36.0803M} Margot, J.~L., Brown, 
M.~E., Trujillo, C.~A., \& Sari, R.\ 2004, AAS/Division for Planetary 
Sciences Meeting Abstracts, 36,  08.03

\bibitem[Margot et al.(2005)]{2005DPS....37.5204M} Margot, J.~L., Brown, 
M.~E., Trujillo, C.~A., Sari, R., \& Stansberry, J.~A.\ 2005, AAS/Division 
for Planetary Sciences Meeting Abstracts, 37,  52.04

\bibitem[Merline et al.(1999a)]{1999IAUC.7129....1M} Merline, W.~J., et al.\ 
1999a, IAUC, 7129

\bibitem[Merline et al.(1999b)]{Me99b} Merline, W.~J., et al.\ 
1999b, Nature, 401, 565  

\bibitem[Merline et al.(2000a)]{Me00a} Merline, W.~J., Close, 
L.~M., Shelton, J.~C., Dumas, C., Menard, F., Chapman, C.~R., \& Slater, 
D.~C.\ 2000a, IAUC, 7503  

\bibitem[Merline et al.(2000b)]{Me00b} Merline, W.~J., Close, 
L.~M., Dumas, C., Shelton, J.~C., Menard, F., Chapman, C.~R., \& Slater, 
D.~C.\ 2000b, Bulletin of the American Astronomical Society, 32, 1017 

\bibitem[Merline et al.(2001a)]{Me01a} Merline, W.~J., Menard, 
F., Close, L., Dumas, C., Chapman, C.~R., \& Slater, D.~C.\ 2001a, IAUC,
7703   

\bibitem[Merline et al.(2001b)]{Me01b} Merline, W.~J., et al.\ 
2001b, IAUC, 7741  

 
\bibitem[Merline et al.(2002)]{2002aste.conf..289M} Merline, W.~J., 
Weidenschilling, S.~J., Durda, D.~D., Margot, J.~L., Pravec, P., \& Storrs, 
A.~D.\ 2002, Asteroids III, 289 

\bibitem[Merline et al.(2002a)]{Me02a} Merline, W.~J., et al.\ 
2002a, IAUC, 7827  

\bibitem[Merline et al.(2002b)]{Me02b} Merline, W.~J., et al.\ 
2002b, IAUC, 7980  
\bibitem[Merline et al.(2003a)]{Me03a} Merline, W.~J., et al.\ 
2003a, IAUC, 8075  

\bibitem[Merline et al.(2003b)]{Me03b} Merline, W.~J., et al.\ 
2003b, IAUC, 8165 

\bibitem[Merline et al.(2003c)]{Me03c} Merline, W.~J., 
Tamblyn, P.~M., Dumas, C., Close, L.~M., Chapman, C.~R., \& Menard, F.\ 
2003,IAUC, 8183  

\bibitem[Merline et al.(2003d)]{Me03d} Merline, W.~J., et al.\ 
2003d,IAUC, 8232 

\bibitem[Merline et al.(2004a)]{Me04a} Merline, W.~J., 
Tamblyn, P.~M., Dumas, C., Menard, F., Close, L.~M., Chapman, C.~R., 
Duvert, G., \& Ageorges, N.\ 2004a, IAUC, 8297 

\bibitem[Merline et al.(2004b)]{2004DPS....36.4601M} Merline, W.~J., et al.\ 
2004b, AAS/Division for Planetary Sciences Meeting Abstracts, 36,  46.01
 
\bibitem[Merline et al.(2005)]{2005DPS....37.0307M} Merline, W.~J., et al.\ 
2005, AAS/Division for Planetary Sciences Meeting Abstracts, 37,  03.07

\bibitem[Mottola et al.(1997)]{Mo97} Mottola, S., Hahn, G., Pravec, P., \& 
Sarounova, L.\ 1997, IAUC, 6680 


\bibitem[Mottola \& Lahulla(2000)]{ML00} Mottola, S., \& 
Lahulla, F.\ 2000, Icarus, 146, 556  

\bibitem[Neish et al.(2003)]{Ne03} Neish, C.~D., Nolan, 
M.~C., Howell, E.~S., \& Rivkin, A.~S.\ 2003, American Astronomical Society 
Meeting Abstracts, 203,  134.02  

\bibitem[Nolan et al.(2000)]{No00} Nolan, M.~C., Margot, 
J.-L., Howell, E.~S., Benner, L.~A.~M., Ostro, S.~J., Jurgens, R.~F., 
Giorgini, J.~D., \& Campbell, D.~B.\ 2000, IAUC, 7518 

\bibitem[Nolan et al.(2002a)]{N02a} Nolan, M.~C., et al.\ 
2002a, IAUC, 7824 

\bibitem[Nolan et al.(2002b)]{N02b} Nolan, M.~C., Howell, 
E.~S., Ostro, S.~J., Benner, L.~A.~M., Giorgini, J.~D., Margot, J.-L., \& 
Campbell, D.~B.\ 2002b, IAUC, 7921  

\bibitem[Nolan et al.(2003a)]{No03a} Nolan, M.~C., Howell, 
E.~S., Rivkin, A.~S., \& Neish, C.~D.\ 2003a, IAUC, 8163 
 
\bibitem[Nolan et al.(2003b)]{N03b} Nolan, M.~C., Hine, 
A.~A., Howell, E.~S., Benner, L.~A.~M., \& Giorgini, J.~D.\ 2003b, IAUC, 
8220 

\bibitem[Nolan et al.(2004a)]{N04a} Nolan, M.~C., Howell, 
E.~S., \& Hine, A.~A.\ 2004a, IAUC, 8336 

\bibitem[Nolan et al.(2004b)]{No04b} Nolan, M.~C., Howell, 
E.~S., \& Miranda, G.\ 2004b, AAS/Division for Planetary Sciences Meeting 
Abstracts, 36,  28.08 

\bibitem[Noll et al.(2002a)]{2002AJ....124.3424N} Noll, K.~S., et al.\ 2002a, 
AJ, 124, 3424 

\bibitem[Noll et al.(2002b)]{N02a} Noll, K., et al.\ 2002b, 
IAUC, 7857 
 
\bibitem[Noll(2003)]{2003EM&P...92..395N} Noll, K.~S.\ 2003, Earth Moon and 
Planets, 92, 395 

\bibitem[Noll et al.(2004a)]{2004AJ....128.2547N} Noll, K.~S., Stephens, 
D.~C., Grundy, W.~M., Osip, D.~J., \& Griffin, I.\ 2004a, AJ, 128, 2547 

\bibitem[Noll et al.(2004b)]{2004Icar..172..402N} Noll, K.~S., Stephens, 
D.~C., Grundy, W.~M., \& Griffin, I.\ 2004b, Icarus, 172, 402 

\bibitem[Ostro et al.(1990)]{1990Sci...248.1523O} Ostro, S.~J., Chandler, 
J.~F., Hine, A.~A., Rosema, K.~D., Shapiro, I.~I., \& Yeomans, D.~K.\ 1990, 
Science, 248, 1523 

\bibitem[Ostro et al.(2000)]{2000IAUC.7496....2O} Ostro, S.~J., Margot, 
J.-L., Nolan, M.~C., Benner, L.~A.~M., Jurgens, R.~F., \& Giorgini, J.~D.\ 
2000, IAUC, 7496 

\bibitem[Ostro et al.(2003)]{O03} Ostro, S.~J., Nolan, 
M.~C., Benner, L.~A.~M., Giorgini, J.~D., Margot, J.~L., \& Magri, C.\ 
2003, IAUC, 8237 

\bibitem[Petit \& Mousis(2004)]{2004Icar..168..409P} Petit, J.-M., \& 
Mousis, O.\ 2004, Icarus, 168, 409 


\bibitem[Pravec \& Hahn(1997)]{1997Icar..127..431P} Pravec, P., \& Hahn, 
G.\ 1997, Icarus, 127, 431 

\bibitem[Pravec et al.(1998)]{1998Icar..133...79P} Pravec, P., Wolf, M., \& 
Sarounova, L.\ 1998, Icarus, 133, 79 

\bibitem[Pravec et al.(2000a)]{2000IAUC.7504....3P} Pravec, P., Kusnirak, 
P., Hicks, M., Holliday, B., \& Warner, B.\ 2000a, IAUC, 7504  

\bibitem[Pravec et al.(2000b)]{P00} Pravec, P., et al.\ 
2000b, Icarus, 146, 190 

\bibitem[Pravec et al.(2001)]{P01} Pravec, P., Kusnirak, 
P., \& Warner, B.\ 2001, IAUC, 7742 

\bibitem[Pravec \& Sarounova(2001)]{PS01} Pravec, P., \& 
Sarounova, L.\ 2001, IAUC, 7633  

\bibitem[Pravec et al.(2002)]{P02} Pravec, P., {\v 
S}arounov{\'a}, L., Hicks, M.~D., Rabinowitz, D.~L., Wolf, M., Scheirich, 
P., \& Krugly, Y.~N.\ 2002, Icarus, 158, 276 

\bibitem[Pravec et al.(2003a)]{P03a} Pravec, P., et al.\ 
2003a, IAUC, 8216 
 
\bibitem[Pravec et al.(2003b)]{P03b} Pravec, P., et al.\ 
2003b, IAUC, 8244 

\bibitem[Pravec et al.(2004)]{P04} Pravec, P., Kusnirak, 
P., Sarounova, L., Brown, P., Kaiser, N., Masi, G., \& Mallia, F.\ 2004, 
IAUC, 8316 

\bibitem[Pravec et al.(2005a)]{P05a} Pravec, P., et al.\ 2005a, Icarus, in press

\bibitem[Pravec et al.(2005b)]{P05b} Pravec, P., Kusnirak, 
P., Kornos, L. Vigli, J.,  Pray, D., Durkee, R., Cooney, W., Gross, J., \& Terrell, D.  \ 2005b, 
IAUC, 8609  

\bibitem[Reddy et al.(2005)]{Rd05} Reddy, V., Dyvig, R., 
Pravec, P., \& Kusnirak, P.\ 2005, IAUC, 8483 

\bibitem[Richardson et al.(1998)]{1998Icar..134...47R} Richardson, D.~C., 
Bottke, W.~F., \& Love, S.~G.\ 1998, Icarus, 134, 47 

\bibitem[Richardson \& Walsh (2005)]{areps05} Richardson, D.~C.,  \& 
Walsh,K.~J.\ 2005, Ann.~Rev.~Earth \& Planet.~Sci., in press

\bibitem[Ryan et al.(2003)]{Ry03} Ryan, W.~H., Ryan, E.~V., 
Martinez, C.~T., \& Stewart, L.\ 2003, IAUC, 8128

\bibitem[Ryan et al.(2004a)]{2004P&SS...52.1093R} Ryan, W.~H., Ryan, E.~V., 
\& Martinez, C.~T.\ 2004a, Planet. Space Sci., 52, 1093 

\bibitem[Ryan et al.(2004b)]{2004DPS....36.4609R} Ryan, W.~H., Ryan, E.~V., 
\& Martinez, C.~T.\ 2004b, AAS/Division for Planetary Sciences Meeting 
Abstracts, 36,  46.09

\bibitem[Schaller \& Brown(2003)]{2003DPS....35.3920S} Schaller, E.~L., \& 
Brown, M.~E.\ 2003, AAS/Division for Planetary Sciences Meeting Abstracts, 
35,  39.20

\bibitem[Schlieder et al.(2004)]{Sl04} Schlieder, J.~E., 
Shepard, M.~K., Nolan, M., Benner, L.~A.~M., Ostro, S.~J., Giorgini, J.~D., 
\& Margot, J.~L.\ 2004, AAS/Division for Planetary Sciences Meeting 
Abstracts, 36,  32.30 

\bibitem[Shepard et al.(2004)]{Sh04} Shepard, M.~K., 
Schlieder, J., Nolan, M.~C., Hine, A.~A., Benner, L.~A.~M., Ostro, S.~J., 
\& Giorgini, J.~D.\ 2004, IAUC, 8397 

\bibitem[Stansberry et al.(2005)]{2005DPS....37.5205S} Stansberry, J.~A., 
Cruikshank, D.~P., Grundy, W.~G., Margot, J.~L., Emery, J.~P., Fernandez, 
Y.~R., \& Rieke, G.~H.\ 2005, AAS/Division for Planetary Sciences Meeting 
Abstracts, 37,  52.05  

\bibitem[Stern(2002)]{2002AJ....124.2300S} Stern, S.~A.\ 2002, AJ, 124, 
2300 

\bibitem[Stern(2003)]{2003LPI....34.1106S} Stern, S.~A.\ 2003, Lunar and 
Planetary Institute Conference Abstracts, 34, 1106 

\bibitem[Stephens et al.(2004)]{2004IAUC.8289....1S} Stephens, D.~C., Noll, 
K.~S., \& Grundy, W.\ 2004, IAUC, 8289

\bibitem[Storrs et al.(2001a)]{St01a} Storrs, A., Vilas, F., 
Landis, R., Wells, E., Woods, C., Zellner, B., \& Gaffey, M.\ 2001a, 
IAUC, 7590  
 
\bibitem[Storrs et al.(2001b)]{St01b} Storrs, A., Vilas, F., 
Landis, R., Wells, E., Woods, C., Zellner, B., \& Gaffey, M.\ 2001b, 
IAUC, 7599  

\bibitem[Tamblyn et al.(2004)]{Tm04} Tamblyn, P.~M., et al.\ 
2004, IAUC, 8293 

\bibitem[Tholen \& Buie(1997)]{TB97} Tholen, D.~J., \& Buie, 
M.~W.\ 1997, Pluto and Charon, eds. S.~A.~Stern \& D.~J.~Tholen, (U.~Arizona Press:Tucson), 193 
 
\bibitem[Veillet et al.(2002)]{2002Natur.416..711V} Veillet, C., et al.\ 
2002, Nature, 416, 711 

\bibitem[Walsh \& Richardson(2005)]{2005DPS....37.1411W} Walsh, K.~J., \& 
Richardson, D.~C.\ 2005, AAS/Division for Planetary Sciences Meeting 
Abstracts, 37,  14.11  

\bibitem[Warner et al.(2005a)]{W05a} Warner, B., Pravec, P., 
Kusnirak, P., Pray, D., Galad, A., Gajdos, S., Brown, P., \& Krzeminski, 
Z.\ 2005a, IAUC, 8511  

\bibitem[Warner et al.(2005b)]{W05b} Warner, B.~D., Pravec, 
P., \& Pray, D.\ 2005b, IAUC, 8592 

\bibitem[Warner et al.(2005c)]{W05c} Warner, B, et al.\ 2005, IAU 
Symp. 229, abstract  

\bibitem[Weaver et al.(1995)]{1995Sci...267.1282W} Weaver, H.~A., et al.\ 
1995, Science, 267, 1282 

\bibitem[Weidenschilling et al.(1989)]{1989aste.conf..643W} 
Weidenschilling, S.~J., Paolicchi, P., \& Zappala, V.\ 1989, Asteroids II, 
643 

\bibitem[Weidenschilling(2002)]{2002Icar..160..212W} Weidenschilling, 
S.~J.\ 2002, Icarus, 160, 212 



\end{thebibliography}
\end{document}